\documentclass[usenatbib]{mnras}
\usepackage{xcolor}

\usepackage{graphicx}
\graphicspath{{./}{./figures/}}

\newcommand*{\mysub}[2]{\ensuremath{#1_{\mathrm{#2}}}}
\newcommand*{\unit}[1]{\ensuremath{\mathrm{\, #1}}}

\newcommand*{\E}[1]{\ensuremath{\times 10^{#1}}}
\newcommand*{\ltsim}{\ {\raise-.75ex\hbox{$\buildrel<\over\sim$}}\ }
\newcommand*{\gtsim}{\ {\raise-.75ex\hbox{$\buildrel>\over\sim$}}\ }

\newcommand*{\Msun}{\ensuremath{\, M_{\odot}}}
\newcommand*{\Zsun}{\ensuremath{\,Z_{\odot}}}
\newcommand*{\mproton}{\mysub{m}{p}}
\newcommand*{\keV}{\unit{keV}}
\newcommand*{\erg}{\unit{erg}}
\newcommand*{\cm}{\unit{cm}}
\newcommand*{\km}{\unit{km}}
\newcommand*{\Mpc}{\unit{Mpc}}
\newcommand*{\second}{\unit{s}}

\newcommand*{\Omegam}{\mysub{\Omega}{m}}
\newcommand*{\rhocr}{\mysub{\rho}{cr}}

\newcommand*{\Mgas}{\mysub{M}{gas}}
\newcommand*{\fgas}{\mysub{f}{gas}}

\newcommand*{\Chandra}{{\it Chandra}}
\newcommand*{\cl}{SPT\,J0459}
\newcommand*{\clfull}{SPT-CL\,J0459$-$4947}

\begin{document}

\title[XMM Observations of the Most Distant SPT Cluster]{Deep XMM-Newton Observations of the Most Distant SPT-SZ Galaxy Cluster}

\author[A.\ B.\ Mantz et al.]{Adam B.\ Mantz,$^{1,2}$\thanks{Corresponding author e-mail: \href{mailto:amantz@stanford.edu}{\tt amantz@stanford.edu}}
  Steven W. Allen,$^{1,2,3}$
  R. Glenn Morris,$^{1,2,3}$\newauthor
  Rebecca E. A. Canning,$^{1,2}$
  Matthew Bayliss,$^{4}$
  Lindsey E. Bleem,$^{5,6}$\newauthor
  Benjamin T. Floyd$^{7}$,
  Michael McDonald$^{8}$
  \smallskip
  \\$^1$Kavli Institute for Particle Astrophysics and Cosmology, Stanford University, 452 Lomita Mall, Stanford, CA 94305, USA
  \\$^2$Department of Physics, Stanford University, 382 Via Pueblo Mall, Stanford, CA 94305, USA
  \\$^3$SLAC National Accelerator Laboratory, 2575 Sand Hill Road, Menlo Park, CA  94025, USA
  \\$^4$Department of Physics, University of Cincinnati, Cincinnati, OH 45221, USA 
  \\$^5$HEP Division, Argonne National Laboratory, Argonne, IL 60439, USA
  \\$^6$Kavli Institute for Cosmological Physics, University of Chicago, Chicago, IL 60637, USA
  \\$^7$Department of Physics and Astronomy, University of Missouri, 5110 Rockhill Road, Kansas City, MO 64110, USA
  \\$^8$Kavli Institute for Astrophysics and Space Research, Massachusetts Institute of Technology, 77 Massachusetts Avenue, \\Cambridge, MA 02139, USA
}
\date{Submitted 22 April 2020, accepted 1 June 2020.}

\maketitle

\begin{abstract}
  We present results from a 577\,ks XMM-{\it Newton} observation of \clfull{}, the most distant cluster detected in the South Pole Telescope 2500 square degree (SPT-SZ) survey, and currently the most distant cluster discovered through its Sunyaev-Zel'dovich effect.
  The data confirm the cluster's high redshift, $z=1.71\pm0.02$, in agreement with earlier, less precise optical/IR photometric estimates.
  From the gas density profile, we estimate a characteristic mass of $M_{500}=(1.8\pm0.2)\E{14}\Msun$; cluster emission is detected above the background to a radius of $\sim2.2\,r_{500}$, or approximately the virial radius.
  The intracluster gas is characterized by an emission-weighted average temperature of $7.2\pm0.3$\,keV and metallicity with respect to Solar of $Z/\Zsun=0.37\pm0.08$.
  For the first time at such high redshift, this deep data set provides a measurement of metallicity outside the cluster center; at radii $r>0.3\,r_{500}$, we find $Z/\Zsun=0.33\pm0.17$, in good agreement with precise measurements at similar radii in the most nearby clusters, supporting an early enrichment scenario in which the bulk of the cluster gas is enriched to a universal metallicity prior to cluster formation, with little to no evolution thereafter.
  The leverage provided by the high redshift of this cluster tightens by a factor of 2 constraints on evolving metallicity models, when combined with previous measurements at lower redshifts.
\end{abstract}


\section{Introduction}

Over the past three decades, sky surveys have progressively extended the discovery space for clusters of galaxies in the Universe.
At present, surveys at X-ray, optical/IR and mm wavelengths -- respectively sensitive to bremsstrahlung emission from the intracluster medium (ICM), galaxy light, and the Sunyaev-Zel'dovich (SZ) effect of the ICM -- have together revealed thousands of massive clusters out to a redshift of $z\sim2$ (e.g.\ \citealt{Ebeling0003191, Ebeling0009101, Bohringer0405546, Rykoff1104.2089, Hasselfield1301.0816, Willis1212.4185, Bleem1409.0850, Planck1502.01598, Adami1810.03849, Gonzalez1809.06820}).
Though still less numerous than those at $z<1$, cluster discoveries at high redshifts are now routine.
The next generation of surveys is poised to extend these searches to redshifts of 3--4, into the epoch where massive, virialized systems first formed.

While the population of massive clusters has now been extensively characterized at $z\ltsim0.5$ (e.g.\ \citealt*{Allen1103.4829}; \citealt{Giodini1305.3286}), with multiwavelength data sets becoming increasingly complete out to $z\sim1$--1.5, few detailed studies have been performed at the highest redshifts now available, $1.5\ltsim z\ltsim2$ (e.g.\ \citealt{Gobat1011.1837, Gobat1907.10985, Andreon1311.4361, Mantz1401.2087, Mantz1703.08221, Brodwin1504.01397, Willis2001.00549}).
Known systems in this range represent a mix of genuinely virialized clusters and forming protoclusters, and, due to their large distances, remain challenging to observe.
Nevertheless, characterizing this population is a key step towards extending our understanding of cluster physics and evolution into the epoch where they first form.

Key questions that observations of these high-redshift clusters can address include the mechanism by which the ICM is enriched with metals, and the impact of non-gravitational processes on the ICM thermodynamics.
The deep potential wells of clusters mean that the ICM retains most of the metals produced by stars in cluster member galaxies.
Observations over the last decade have revealed a universal and spatially uniform enrichment of the ICM in low-redshift clusters (e.g.\ \citealt{Werner1310.7948, Simionescu1506.06164, Simionescu1704.01236}).
This implies early enrichment to a constant level in the protocluster intergalactic medium, prior to cluster formation, at $z\gtsim2$, and makes the firm prediction that clusters at redshifts 1.5--2 should have similar metal abundances to nearby clusters.

In this work, we have targeted the highest-redshift cluster discovered in the South Pole Telescope (SPT) 2500 square degree SZ survey with detection significance $>5$, \clfull{} (henceforth \cl{}).
The original SPT-SZ cluster catalog contained 3 clusters whose redshifts were conservatively reported as $z>1.5$, since the available ground-based optical and {\it Spitzer} IR photometric data could not provide a precise redshift estimate beyond this limit \citep{Bleem1409.0850}.
Of these, \cl{} has the highest SZ detection significance and is thus presumably the most massive.
Of the others, SPT-CL\,J0446$-$4606 now has a revised photometric redshift of $1.52^{+0.13}_{-0.02}$ from {\it Hubble} and {\it Spitzer} data \citep{Strazzullo1807.09768}, while SPT-CL\,J0334$-$4645 was detected with survey significance $<5$ and therefore not prioritized for confirmation (nor used in, e.g., cosmological analyses; \citealt{de-Haan1603.06522}).
\cl{} is thus the highest-redshift confirmed cluster found by SPT-SZ, and, at this writing, is the highest-redshift cluster originally discovered through its SZ effect.
With the results reported here, it joins a small group of confirmed, massive clusters at $z>1.7$, the others having all been discovered using X-ray or IR data.

\Chandra{} observations of \cl{} were able to measure the cluster's X-ray luminosity and ICM density \citep{McDonald1702.05094}.
Based on its known flux, we designed an XMM-{\it Newton} observing program aimed at resolving the metallicity and temperature structure of the cluster in multiple radial ranges, a first for such a high redshift.
While exceedingly modest compared with studies that have routinely been done at low redshift, these goals required several hundred ks of observing time, even with XMM's high throughput.
We describe the reduction of these data, along with the archival \Chandra{} data, in Section~\ref{sec:data}.
Section~\ref{sec:model} covers our modeling of the ICM and contaminating point sources, as well as the XMM point spread function (PSF), which must be accounted for to study this distant, compact source.
Our results and their implications are presented in Sections~\ref{sec:results} and \ref{sec:discussion}. We conclude in Section~\ref{sec:conclusion}.

All cosmology-dependent quantities reported in this work were computed for a reference flat, cosmological-constant model with Hubble parameter $70\km\second^{-1}\Mpc^{-1}$ and matter density with respect to the critical density $\Omegam=0.3$.
Parameter values correspond to marginalized posterior modes, and quoted uncertainties to 68.3 per cent highest posterior density credible intervals.
The characteristic mass and radius of a cluster are jointly defined by the relation $M_\Delta = (4/3)\pi\,\Delta\,\rhocr(z) r_\Delta^3$, where $\rhocr(z)$ is the critical density at the cluster's redshift, and $\Delta=500$.

\section{Data Reduction} \label{sec:data}

\subsection{XMM-Newton} \label{sec:data_xmm}

\cl{} was observed by XMM for a total of 577\,ks (before filtering) between April and November, 2017 (Table~\ref{tab:observations}).
These data were reduced using the XMM-{\it Newton} Extended Source Analysis Software ({\sc xmm-esas}; {\sc sas} version 18.0.0),\footnote{\url{http://www.cosmos.esa.int/web/xmm-newton/sas}} following the recommendations of \citet{Snowden0710.2241} and the {\sc xmm-esas} Cookbook.\footnote{\url{http://heasarc.gsfc.nasa.gov/docs/xmm/esas/cookbook/}}
Following standard calibration and filtering of the raw event files, lightcurves for each of the EPIC detectors were manually inspected, and periods in which the X-ray background was enhanced (flared) were manually removed.
The clean exposure times for the MOS (PN) detectors, after accounting for overheads and lightcurve filtering, were 56 (66) per cent of the total, in line with historical expectations \citep{Salvetti1705.04172}.
The final clean exposure times for the PN, MOS1 and MOS2 are, respectively, 378, 379 and 321\,ks.
From these event files, we produced images, quiescent background maps and exposure maps corresponding to 0.4--4.0\,keV observer-frame energies for our imaging analysis (Section~\ref{sec:model_psf}), as well as spectra, responce matrices and ancillary response files for our later spectral analysis.

\begin{table}
  \centering
  \caption{
    Clean exposure times after automatic and manual filtering for each of the XMM (MOS1, MOS2 and PN) and \Chandra{} (ACIS-I) detectors used in this work.
  }
  \vspace{1ex}
  \begin{tabular}{crrr@{\hspace{8mm}}cr}
    OBSID & MOS1 & MOS2 & PN & OBSID & ACIS-I \\
    \hline
    0801950101 & 101 & 103 & 89 & 17211 & 13 \\
    0801950201 & 97 &  96 & 85 & 17501 & 22 \\
    0801950301 & 95 & 95 & 82 & 17502 & 14 \\
    0801950401 & 67 & 66 & 52 & 18711 & 23 \\
    0801950501 & 18 & 19 & 14 & 18824 & 22 \\
    &&&& 18853 & 30 \\
    \hline
    Total (ks) & 378 & 379 & 321 & & 123 \\
  \end{tabular}
  \label{tab:observations}
\end{table}

\begin{figure*}
  \centering
  \includegraphics[scale=0.35]{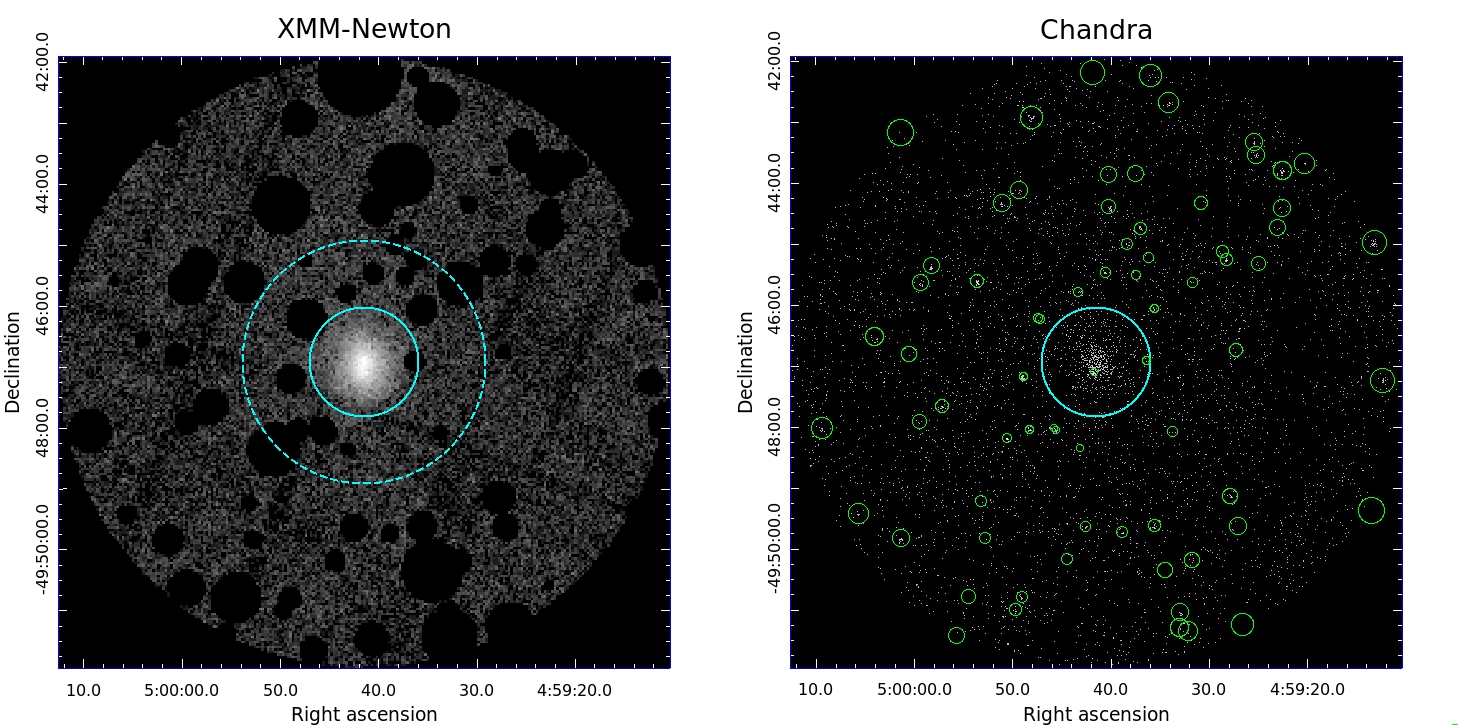}
  \caption{
    Stacked images from the XMM and \Chandra{} observations of \cl{}.
    The solid cyan circle shows our estimate of $r_{500}\approx54''$ (Section~\ref{sec:results_deproj}), while the dashed cyan circle approximately corresponds to the detected extent of the cluster emission (radius $2'$; Section~\ref{sec:model_psf}).
    Point-like sources identified in the \Chandra{} data are circled in green in the right panel (Section~\ref{sec:model_agn}).
  } \label{fig:images}
\end{figure*}

\subsection{Chandra} \label{sec:data_chandra}

We use archival \Chandra{} observations of \cl{}, totaling 139\,ks (see \citealt{McDonald1702.05094}). 
The data were processed as described by \citet{Mantz1502.06020}, with the exception that this work employs a more recent version of the \Chandra{} calibration files (specifically, {\sc caldb}\footnote{\url{http://cxc.harvard.edu/caldb/}} version 4.7.1).
In brief, the raw data were reprocessed to produce level 2 event files, and were filtered to eliminate periods of high background.
The total clean exposure time after this step was 123\,ks.
For each observation, a corresponding quiescent background data set was produced using the \Chandra{} blank-sky data,\footnote{\url{http://cxc.cfa.harvard.edu/ciao/threads/acisbackground/}} rescaled according to the measured count rate in the 9.5--12\,keV band.
Our pipeline produces images, blank-sky background maps and exposure maps in the 0.6--2.0\,keV band.

\section{Modeling} \label{sec:model}

\subsection{Point-like sources} \label{sec:model_agn}

A preliminary catalog of active galactic nuclei (AGN) detected in the \Chandra{} data by the Cluster AGN Topography Survey pipeline (CATS; Canning et~al., in preparation) was used to produce a mask for the \Chandra{} and XMM data.
In the case of XMM, the measured CATS flux of each AGN was used to resize the corresponding circular mask, such that the maximum surface brightness \emph{not} covered by the mask was $5\E{-19}$\,erg\,s$^{-1}$\,cm$^{-2}$\,arcsec$^{-2}$ in the 0.4--4\,keV band used in our image analysis below; this calculation used the XMM PSF model described in the next section.
A maximum radius of $60''$ was enforced in this process, though it was never realized within the $5'$ radius region used in our analysis throughout this work.
While these particular choices were made to produce visually reasonable masks, we note that the residual, unmasked brightness is significantly less than the residual background level found in the next section.
The one exception to this process is for a relatively faint AGN appearing within the bulk of the cluster emission (2--10\,keV flux of $4.3\E{-15}$\,erg\,s$^{-1}$\,cm$^{-2}$, projected $11''$ from the cluster center) which cannot practically be masked in the same way.
Instead, the contaminating flux from this source is forward modeled in our spectral analysis (Section~\ref{sec:spectral}), assuming a power law spectrum with index $-1.4$; the source is faint enough that assuming a steeper index or even neglecting it entirely has a $<1\sigma$ impact on our spectral results.
In addition, a small number of variable sources that are visible in the XMM but not the \Chandra{} data were masked manually.
The stacked XMM and \Chandra{} images, respectively showing the mask and the detected point sources, appear in Figure~\ref{fig:images}.
The final mask used in the XMM analysis removes 20 per cent of the area in the central $5'$ radius circle.

\subsection{XMM-Newton point spread function} \label{sec:model_psf}

\begin{figure*}
  \centering
  \includegraphics{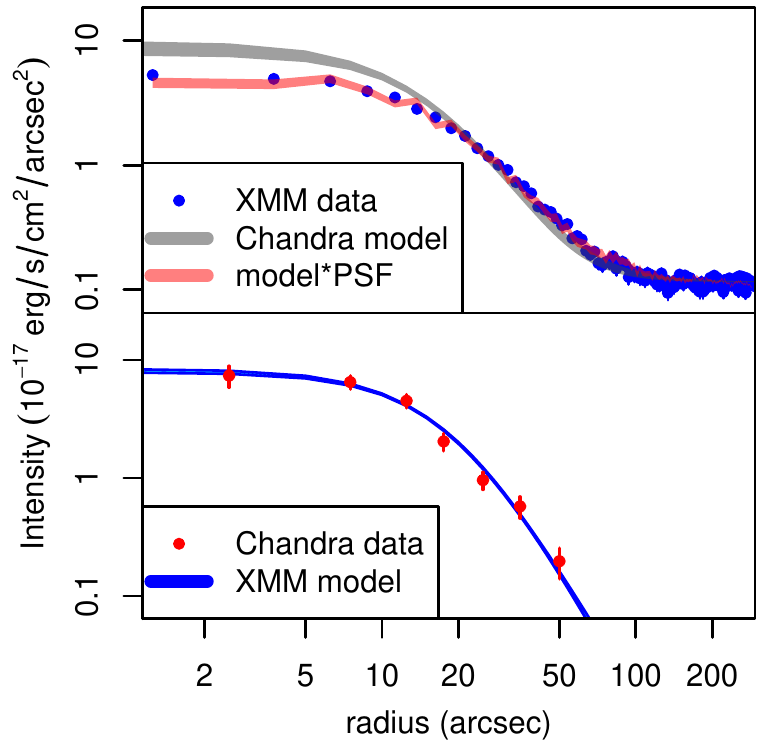}
  \hspace{10mm}
  \includegraphics{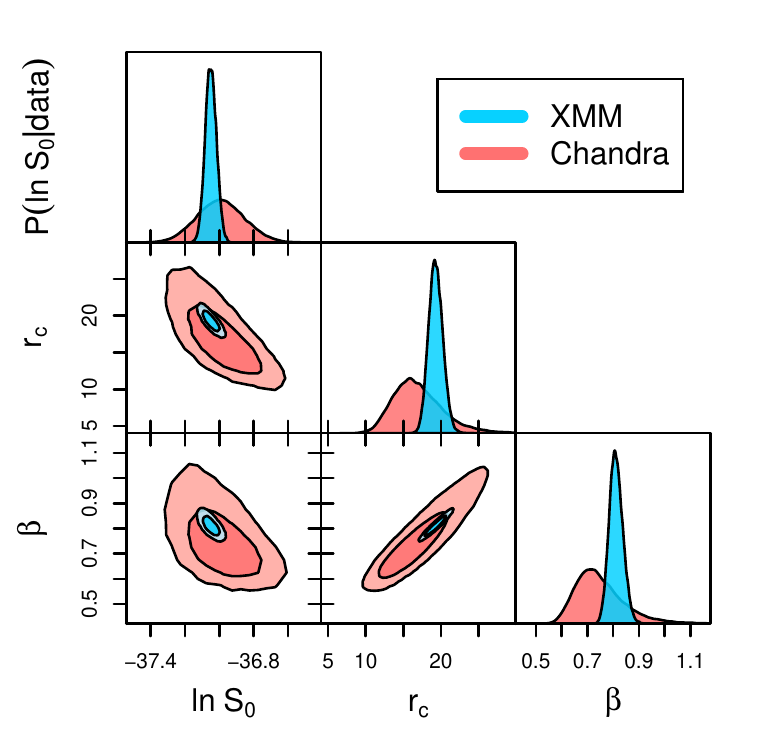}
  \caption{
    Left: intensity profiles of \cl{} in the 0.4--4.0\,keV band from XMM and \Chandra{} data.
    Profiles are binned for display, although our analysis runs on profiles made at pixel resolution.
    A $\beta$ model plus constant background is fit to each data set independently.
    In the upper panel, the XMM data are compared with the sum of the $\beta$ model fitted to the \Chandra{} data and the constant background fitted to the XMM data (gray shading).
    The red shading shows this model after being convolved with the XMM PSF.
    Conversely, the bottom panel compares the \Chandra{} profile with the sum of the $\beta$ model from the XMM fit (accounting for the PSF) and the residual \Chandra{} background fit.
    Right: one-dimensional (on-diagonal) and two-dimensional (off-diagonal) marginalized posteriors of the $\beta$ model parameters from XMM (blue) and \Chandra{} (red).
    Here $S_0$ has units of erg$\second^{-1}\cm^{-2}$\,arcsec$^{-2}$ and $\mysub{r}{c}$ has units of arcsec.
    See Section~\ref{sec:model_psf}.
  } \label{fig:sbprof}
\end{figure*}

The high redshift of \cl{} means that our analysis can be confined to the central portion of the FoV and to relatively low energies ($<4$\,keV).
We therefore adopt a model for the XMM PSF that is axisymmetric and monochromatic, consisting of the sum of a Gaussian ``core'' and a more extended $\beta$ profile \citep{Read1108.4835}, and apply it to each EPIC detector.
As a sanity check, we extract surface brightness profiles from the XMM and \Chandra{} data about a common center, converting these to intensity units (erg\,s$^{-1}$\,cm$^{-2}$\,arcsec$^{-2}$)  using {\sc pimms},\footnote{\url{https://heasarc.gsfc.nasa.gov/cgi-bin/Tools/w3pimms/w3pimms.pl}}$^{,}$\footnote{\url{http://cxc.harvard.edu/toolkit/pimms.jsp}} assuming a fiducial temperature of 7\,keV, a metallicity of $0.3\Zsun$ and a redshift of 1.7, and accounting for Galactic absorption (see Section~\ref{sec:model_icm}).
We fit a $\beta$ model plus constant residual background to each data set independently, excluding the point source masks described in the previous section.
For \Chandra{}, we use the ``blank sky'' data sets (rescaled to account for relative exposure times) from standard data reduction, which in principle contain all astrophysical and instrumental backgrounds.
The model is scaled by the exposure map, added to the blank sky image, and compared to the measured counts in the science image using the \citet{Cash1979ApJ...228..939} statistic (i.e., using a Poisson likelihood function).
We used the {\sc rgw}\footnote{\url{https://github.com/abmantz/rgw}} Markov Chain Monte Carlo (MCMC) code to obtain constrains on the model parameters.
For XMM, the procedure is slightly different.
We use the model quiescent particle background maps generated by SAS, which represent only a fraction of the total background, in lieu of ``blank sky'' images; hence the constant background model parameter fits to a positive value in this case rather than being consistent with zero.
More importantly, the $\beta$ model is convolved with the PSF before being compared with the data.
Figure~\ref{fig:sbprof} shows that the cluster model, fit to the XMM data in this way, is consistent with the higher-resolution \Chandra{} surface brightness profile.
Conversely the model fit to the \Chandra{} data, if convolved with the XMM PSF model, matches the XMM profile well.
We have checked that the same conclusions hold when using XMM images in separate soft and (relatively) hard energy bands (0.4--1.5 and 1.5--4.0\,keV).
This agreement gives us confidence that the simple PSF model employed here is adequate for our analysis below.

In the XMM surface brightness profile, cluster emission can be seen (that is, the net enclosed counts in excess of the background are clearly increasing) out to a radius of $\sim2'$.
This is a factor of 2.2 times the estimate of $r_{500}$ we arrive at in Section~\ref{sec:results_deproj}, meaning that the cluster is visible to approximately the virial radius.
This is due to both the depth of our data, the relatively higher physical density at the virial radius at high redshifts compared with low redshifts, and (to a lesser extent) the large angular diameter distance to the cluster leading to an increased surface brightness.
By comparison, XLSSC\,122, a fainter cluster at $z=1.98$, was detected to approximately $r_{200}\approx1.5\,r_{500}$ in 100\,ks of XMM observation \citep{Mantz1703.08221}.

\subsection{Intracluster medium} \label{sec:model_icm}

We adopt (4:59:41.4,$-$49:46:56; J2000 coordinates), the median photon position in the \Chandra{} image after masking point sources \citep{Mantz1502.06020}, as the cluster center.
This is within $1''$ of the bright central galaxy position, and is preferable to using a center determined from the XMM data due to the presence of a faint point source within the bright part of the cluster (Section~\ref{sec:model_agn}).
With the exception of the ``single spectrum'' results in Section~\ref{sec:results_bucket}, our analysis seeks to determine properties of the ICM as a function of radius in three dimensions.
In these cases, spectra are therefore extracted in at least two concentric annuli.
The cluster model consists of emission from a series of concentric spherical shells, with inner and outer three-dimensional radii corresponding to the projected radii of the annuli within which spectra are extracted.
However, a given annulus potentially contains signal from every spherical shell, for two reasons.
First, the emission from each shell is geometrically projected onto all annuli at equal or smaller radius.
Second, the PSF spreads photons originating at a given position on the sky across all annuli at some level.
For a given set of annuli/shells, we compute the overall mixing matrix, encoding the fraction of emission from each shell ending up in each annulus.
Within this calculation, we use the best fitting $\beta$ model fit to the XMM data (converted to a 3-dimensional emissivity profile, when considering the mixing due to geometric projection) as a template for the radial dependence of emissivity within each shell/annulus.
Note that this aspect differs from the geometric projection model {\sc projct} in {\sc xspec}, which assumes a constant emissivity within each shell, and also from the use of ``cross-talk'' ancillary response files to model PSF mixing.
Given the prior that the cluster emission resembles a $\beta$ model, which in this case is directly verifiable using the higher-resolution \Chandra{} data, our approach should result in a more accurate model of the mixing between the cluster center and the outskirts.

\begin{figure*}
  \centering
  \includegraphics{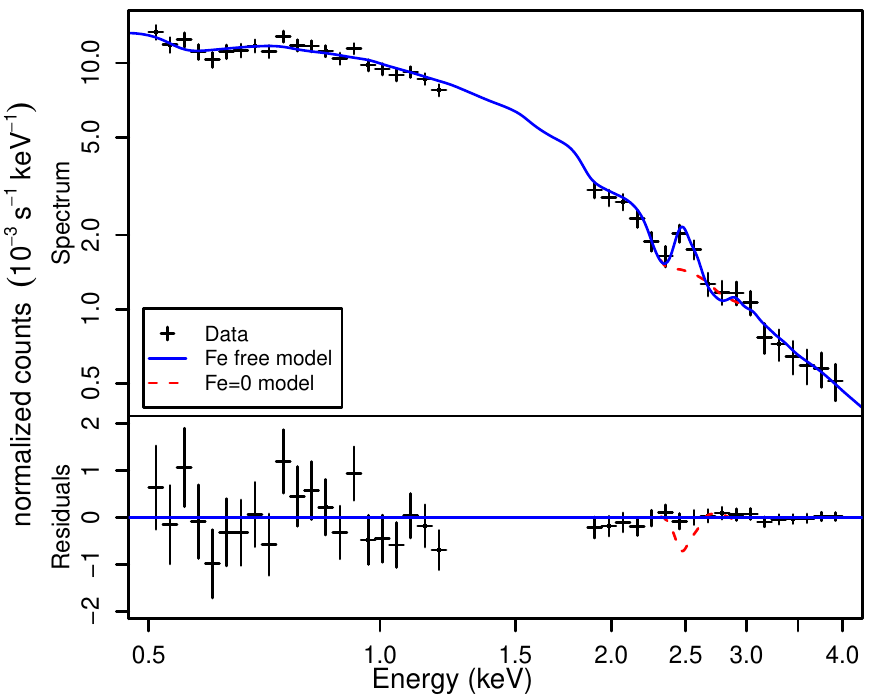}
  \hspace{5mm}
  \includegraphics{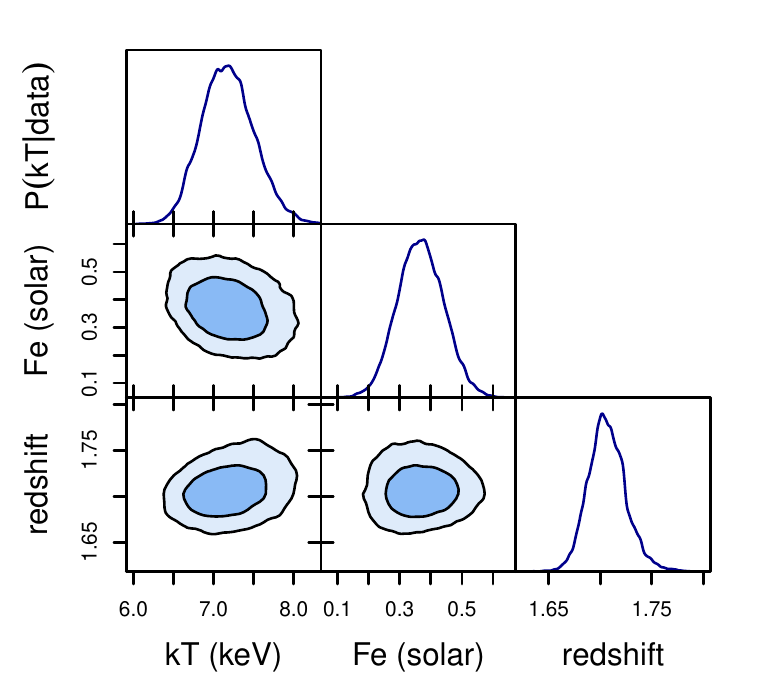}
  \caption{
    Left: stacked and background-subtracted (for display only) EPIC spectrum of the central arcminute of \cl{}.
    Energies in the 1.2--1.9\,keV range are excluded due to contamination from instrumental emission lines.
    The blue, solid curve shows the best-fitting single thermal component model, with a free temperature, metallicity, redshift and normalization.
    The red, dashed curve shows the best fitting model where the metallicity is fixed to zero, to indicate the strength of the Fe emission feature.
    The lower panel shows residuals with respect to the model with fitted metallicity.
    Right: one-dimensional (on-diagonal) and two-dimensional (off-diagonal) marginalized posteriors for the cluster emission model fit to these data.
  }
  \label{fig:spectrum}
\end{figure*}

\subsection{Spectral analysis details} \label{sec:spectral}

Our spectral analysis uses {\sc xspec}\footnote{\url{http://heasarc.gsfc.nasa.gov/docs/xanadu/xspec/}} (version 12.10) to compute spectral models and evaluate likelihoods, and the {\sc lmc} MCMC code\footnote{\url{https://github.com/abmantz/lmc}} to estimate parameter posterior distributions.
Bremsstrahlung continuum and line emission from the ICM were evaluated using the {\sc apec} plasma model (ATOMDB version 3.0.9), parametrized by a temperature, metallicity (with relative abundances of different metals fixed to the Solar ratios of \citealt{Asplund0909.0948}), redshift, and overall normalization.
Photoelectric absorption by Galactic gas was accounted for using the {\sc phabs} model, employing the cross sections of \citet{Balucinska1992ApJ...400..699B}, and adopting a fixed value of $1.19\E{20}\cm^{-2}$ for the equivalent absorbing hydrogen column density \citep{Kalberla0504140}; updating this to the more recent value of $1.08\E{20}\cm^{-2}$ \citep{HI4PI1610.06175} has a negligible impact on our results.
As mentioned in Section~\ref{sec:model_agn}, we model the contribution from a single AGN identified in the \Chandra{} data but not masked.
We model the remaining astrophysical and instrumental backgrounds empirically using a spectrum extracted from an annulus spanning radii $3'$--$5'$ from the cluster center.
Our surface brightness model from Section~\ref{sec:model_psf} indicates that the cluster signal at $3'$ is $\ltsim1$ per cent of the background, and the cluster flux expected to be mixed from the background region to radii $<2'$ (the greatest extent of our cluster analysis below) is correspondingly negligible.
The model parameters are fit using the \citet{Cash1979ApJ...228..939} statistic, as modified for {\sc xspec} by \citet[][the $C$ statistic]{Arnaud1996ASPC..101...17A}, to properly account for the Poisson nature of the source and background counts.
Spectra were binned to have at $\geq1$ count per channel, as the modified $C$ statistic is known to be biased when the data include empty channels.\footnote{{\sc xspec} manual Appendix B: \url{https://heasarc.gsfc.nasa.gov/docs/xanadu/xspec/manual/node304.html}}
We fit data in the observer-frame energy band 0.5--4.0\,keV (rest frame energies $\sim1.4$--11\,keV), with the exception of the ranges 1.2--1.9\,keV (for MOS) and 1.2--1.65\,keV (for PN), which are heavily contaminated by aluminum and (in MOS) silicon instrumental emission lines.

\section{Results} \label{sec:results}

\subsection{Single spectrum} \label{sec:results_bucket}

We first fit XMM spectra extracted from a single circular region of radius $1'$ about the cluster center.
While detectable emission from the cluster extends to $\sim2'$ radius, this smaller region contains $>90$ per cent of the total cluster signal; the radius of $1'$ is also conveniently very close to the estimate of $r_{500}\approx 54''$ derived in the next section.
From these data, we constrain the ``average'' properties of the ICM to be $kT=7.2\pm0.3$, $Z/\Zsun=0.37\pm0.08$ and $z=1.705\pm0.018$.
Our redshift constraint, due to the (rest frame) 6.7\,keV Fe emission line complex, is consistent with earlier photometric estimates of $1.7\pm0.2$ ({\it Spitzer} IR and ground-based NIR data; \citealt{Bleem1409.0850}) and $1.80^{+0.10}_{-0.19}$ ({\it Spitzer} and {\it Hubble} imaging; \citealt{Strazzullo1807.09768}).
Figure~\ref{fig:spectrum} shows the stacked XMM EPIC spectra used in this fit, and the corresponding parameter constraints (see also Table~\ref{tab:spectral}).
Note that we continue to marginalize over the cluster redshift in the spectral fits below, rather than fixing it based on these results.

\begin{figure*}
  \centering
  \begin{minipage}[b]{0.45\textwidth}
    \includegraphics{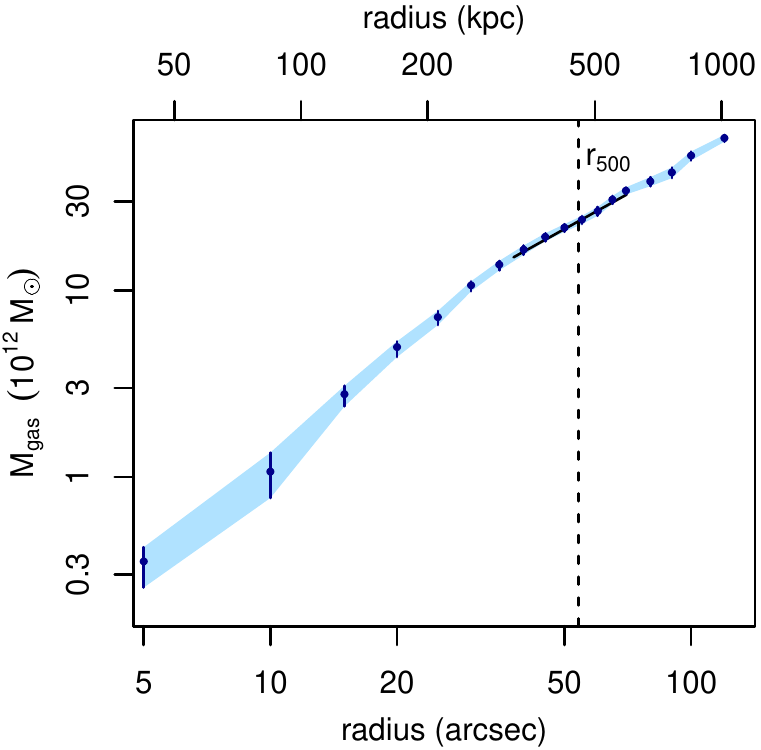}
    \vspace{0.25in}
  \end{minipage}
  \hspace{10mm}
  \includegraphics{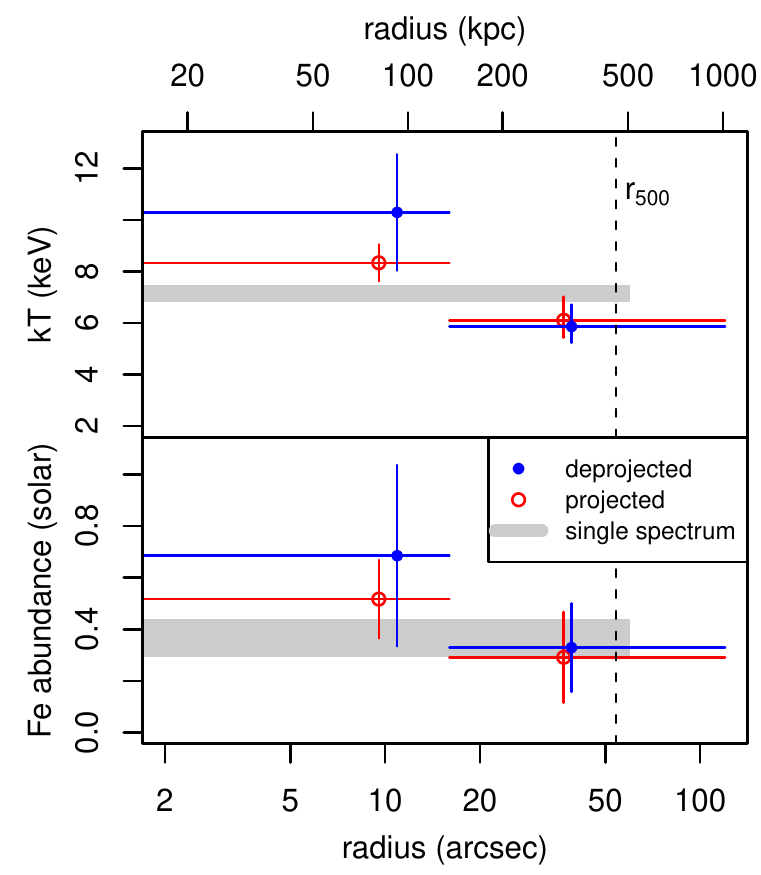}
  \caption{
    Left: enclosed gas mass profile of \cl{} as determined from our analysis in Section~\ref{sec:results_deproj}.
    While the spectral fit leading to this profile marginalizes over cluster redshift, the values of gas mass and radius in physical units shown here assume a cluster redshift of $z=1.71$.
    Error bars and shading indicate the 68.3 per cent confidence intervals as a function of radius.
    The vertical, dashed line shows our estimate of $r_{500}$, which is insensitive to the assumed redshift within the posterior redshift uncertainty.
    The solid line shows the typical power-law slope of the gas mas profile at $\sim r_{500}$, 1.25, measured from $z<0.5$ clusters \citep{Mantz1606.03407}, with which the measured profile of \cl{} agrees well.
    Right: constraints on temperature and metallicity from the single-spectrum analysis of Section~\ref{sec:results_bucket}, as well as the 2-annulus analysis of Section~\ref{sec:results_thermo}.
    Blue points show the results obtained from our standard analysis, accounting for mixing due to both the PSF and geometrical projection (i.e., the values correspond to spherical shells of gas).
    For comparison, red points show the results when geometrical projection is not accounted for, corresponding instead to the projected properties of the two annuli.
    Points are shown at the emission-weighted average radius associated with each measurement, either in projection or in 3D.
  }
  \label{fig:Mgas}
  \label{fig:kTZprof}
\end{figure*}

\begin{table*}
  \centering
  \caption{
    Measured temperatures and metallicities of \cl{}, marginalized over the cluster redshift in all cases.
    Shown are results from a single spectrum; two concentric annuli, where mixing of emission from the PSF is accounted for but geometric projection is not; and the same annuli when both the PSF and projection are accounted for.
    Net (background subtracted) counts are in the 0.5--4.0\,keV band.
    The final column compares the best $C$ statistic with the number of degrees of freedom.
  }
  \label{tab:spectral}
  \begin{tabular}{ccccccccccc}
    \hline\vspace{-3ex}\\
    Deprojected & $r_\mathrm{inner}^{(1)}$ & $r_\mathrm{outer}^{(1)}$ & $kT^{(1)}$ & $Z^{(1)}$ & $r_\mathrm{inner}^{(2)}$ & $r_\mathrm{outer}^{(2)}$ & $kT^{(2)}$ & $Z^{(2)}$ & Net counts & cstat/dof\vspace{1ex}\\
                & $('')$ & $('')$ & (keV) & $(\Zsun)$ & $('')$ & $('')$ & (keV) & $(\Zsun)$\vspace{0.2ex}\\
    \hline\vspace{-2ex}\\
    No & 0 & 60 & $7.2\pm0.3$ & $0.37\pm0.08$ &---&---&---&---& 11499 & 4351/5121\vspace{0.5ex}\\
    No & 0 & 16 & $8.3\pm0.7$ & $0.52\pm0.15$ & 16 & 120 & $6.1^{+0.9}_{-0.7}$ & $0.29\pm0.18$ & 12704 & 8109/8998\vspace{0.5ex}\\
    Yes & 0 & 16 & $10.3\pm2.3$ & $0.69\pm0.35$ & 16 & 120 & $5.9^{+0.9}_{-0.6}$ & $0.33\pm0.17$ & 12704 & 8109/8998\vspace{0.5ex}\\
    \hline
  \end{tabular}
\end{table*} 

\begin{figure*}
  \centering
  \includegraphics{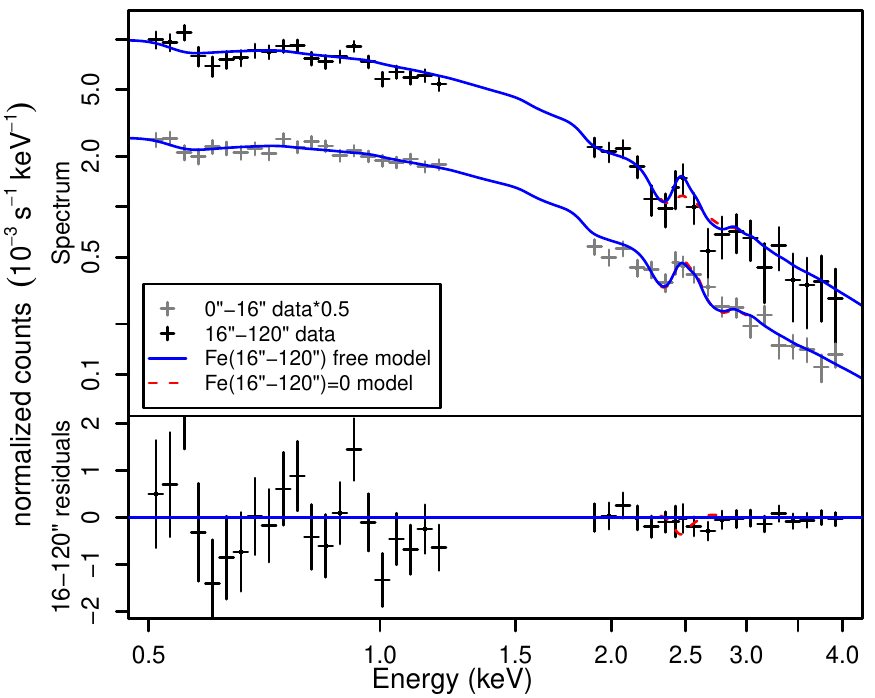}
  \hspace{5mm}
  \includegraphics{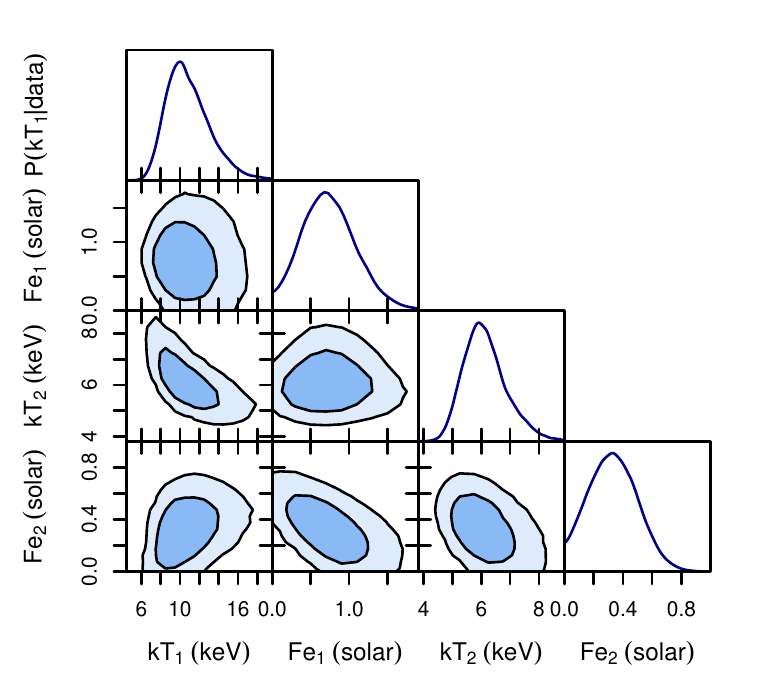}
  \caption{
    As Figure~\ref{fig:spectrum}, but for the deprojection analysis in Section~\ref{sec:results_thermo}, showing the posteriors for the temperature and metallicity of two spherical shells.
    The cluster redshift is marginalized over, although we do not include it in the right panel; constraints are similar to those in Figure~\ref{fig:spectrum}.
    The ``Fe=0'' and ``Fe free'' models in the left panel refer to the metallicity in the outer shell.
    The metallicity of the inner shell is free in both cases, and PSF mixing thus produces an emission line feature in the outer annulus even for the ``Fe=0'' case.
    The data and models for the inner annulus are scaled by a factor of 0.5 in the left panel for clarity.
  }
  \label{fig:deproj}
\end{figure*}

\subsection{Density deprojection} \label{sec:results_deproj}

To constrain the ICM density profile, we fit spectra extracted in annuli whose outer radii correspond to the points shown in the left panel of Figure~\ref{fig:Mgas}.
The largest radius included in the cluster model is $2'$, corresponding to a signal-to-background ratio of $\sim0.05$, according to the surface brightness model of Section~\ref{sec:model_psf}.
Using the methods described in Section~\ref{sec:model}, we fit for the emissivity in each of the corresponding spherical shells, as well as a common temperature, metallicity and redshift.
Figure~\ref{fig:Mgas} shows the resulting constraints on the integrated gas mass profile, where we have assumed a canonical value of the mean molecular mass of $\mu = 0.61\mproton$ and a cluster redshift of $z=1.71$ when converting emissivity to gas density, and ultimately to gas mass, and angular separation to metric radius.

We arrive at an estimate of $M_{500}$, and the corresponding radius $r_{500}$, by solving the implicit equation
\begin{equation} \label{eq:rDelta}
  M(r_{500}) = \frac{\Mgas(r_{500})}{\fgas(r_{500})} = \frac{4\pi}{3} \, 500\rhocr(z) r_{500}^3.
\end{equation}
For this purpose, we adopt a fiducial value of $\fgas(r_{500})=0.125$, based on results from massive, X-ray selected clusters at $z<0.5$ \citep{Mantz1606.03407}, and also consistent with dynamically relaxed clusters at redshifts $z<1.08$ \citep{Mantz1509.01322}.
While this value of $\fgas(r_{500})$ has not been independently verified to apply at $z\sim1.7$, the gas mass fraction is theoretically expected to be non-evolving at intermediate radii for massive clusters (e.g.\ \citealt{Eke9708070}; \citealt*{Nagai0609247}; \citealt{Battaglia1209.4082, Planelles1209.5058, Barnes1607.04569, Singh1911.05751}).
This is supported by the good agreement between the slope of the $\Mgas(r)$ profile at these radii and the average value for the cluster sample of \citet{Mantz1606.03407}, shown in Figure~\ref{fig:Mgas}.
This procedure yields a characteristic radius of $r_{500}=(458\pm18)$\,kpc ($54''\pm2''$) and mass of $M_{500}=(1.8\pm0.2)\E{14}\Msun$, when using a cluster redshift of $z=1.71$ in the unit conversions discussed above; this marginally lower than the mass of $(2.7\pm0.6)\E{14}\Msun$ estimated from the SZ signal by \citet{Bleem1409.0850}.
Accounting for the posterior uncertainty on the cluster redshift in these computations has a negligible impact, at the $<1$ per cent level for $r_{500}$ (as expressed in kpc).

\subsection{Temperature and metallicity deprojection} \label{sec:results_thermo}

To move beyond a single free temperature and metallicity for the ICM, we simplify the analysis to spectra extracted within two annuli, rather than the large number of annuli used in Section~\ref{sec:results_deproj}.
This minimizes the impact of binning each spectrum to $\geq1$ count per channel, and the corresponding loss of energy resolution, on fitting the Fe emission feature.
Table~\ref{tab:spectral} and the right panel of Figure ~\ref{fig:kTZprof} shows the temperature and metallicity profile constraints at radii of $0''$--$16''$ and $16''$--$2'$, corresponding to 0.0--0.3\,$r_{500}$ and 0.3--2.2\,$r_{500}$.
In addition to these ``deprojected'' results, showing constraints on these quantities in spherical shells, the figure compares to constraints where the PSF, but not geometrical projection, is accounted for, as well as the single-spectrum results from Section~\ref{sec:results_bucket}.
Points in the right panel of Figure ~\ref{fig:kTZprof} are located at the emission-weighted average radius associated with each annulus or shell, approximately reflecting which radii are most influential in the fit.

\section{Discussion} \label{sec:discussion}

\subsection{ICM metallicity evolution}

\begin{figure}
  \centering
  \includegraphics{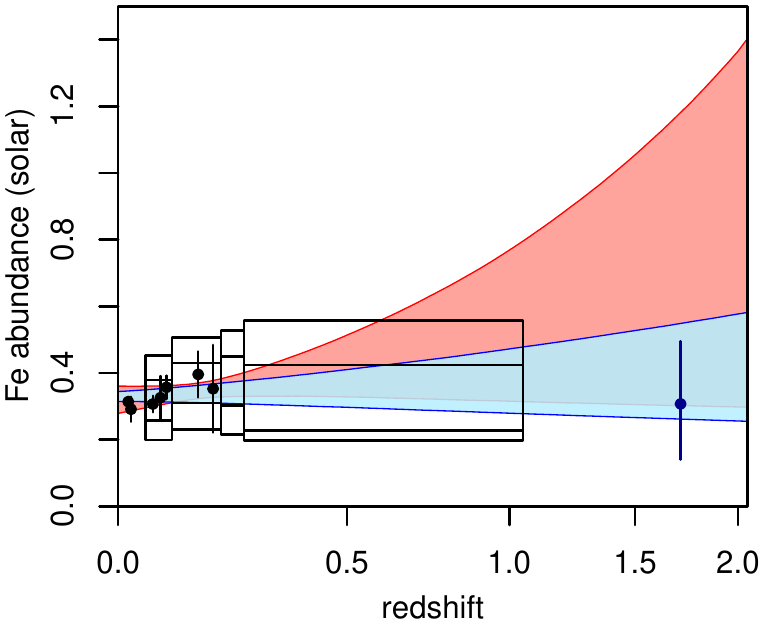}
  \caption{
    Measured center-excluded ICM metallicities from {\it Suzaku} and \Chandra{}, and our observations of \cl{}, and constraints on evolution models.
    Black points show {\it Suzaku} measurements of low-redshift clusters \citep{Werner1310.7948, Simionescu1302.4140, Urban1706.01567}, while black boxes represent the 68.3 and 95.4 per cent limits from clusters observed by \Chandra{} in different redshift ranges \citep{Mantz1706.01476}.
    Red shading shows the 68.3 per cent confidence limits for power-law models, $\propto(1+z)^\gamma$, from these data.
    The blue point at $z=1.71$ represents our measurement of \cl{}, and light blue shading shows the improved model constraints when it is combined with the {\it Suzaku} and \Chandra{} data.
  } \label{fig:metalz}
\end{figure}

Multiple lines of evidence point to a uniform and universal enrichment of the intergalactic medium with the products of stellar evolution at high redshifts ($z\gtsim2$) and prior to the formation of massive clusters of galaxies.
In the brightest, nearby clusters, where abundances of Fe and other heavy elements can be measured with high spatial resolution (in metric units) and signal-to-noise, spatially uniform metal distributions have been observed outside of cluster cores, extending even to their virial radii \citep{Werner1310.7948, Simionescu1506.06164, Simionescu1704.01236}.
Azimuthally averaged metallicity profiles have shown a central peak, whose magnitude depends on the presence or absence of a cool core, declining to a universal value  ($\sim0.3\Zsun$ for Fe) at larger radii (e.g.\ \citealt{Allen9802219, De-Grandi0310828, Leccardi0806.1445, Tholken1603.05255, Mernier1703.01183, Urban1706.01567, Urdampilleta1906.08067}).
Studies of the evolution of the ICM metallicity have produced a variety of conclusions, partly due to differences in the radial ranges probed by different observations (e.g.\ \citealt{Balestra0609664, Leccardi0806.1445, Maughan0703156, Anderson0904.1007, Andreon1209.0565, Baldi1111.4337, Ettori1504.02107, McDonald1603.03035, Mantz1706.01476}).
The picture that emerges is one where the gas that forms the ICM was pre-enriched and well mixed to a universal level prior to the formation of a deep potential well, with stellar evolution, galactic winds, and redistribution of gas by AGN feedback continuing to slowly enrich and mix the ICM near cluster centers, while the universal metallicity value is retained at larger radii.
Simulations support this notion, finding that AGN feedback at $z\gtsim2$, when cluster potentials are still relatively shallow, is particularly effective at mixing metals throughout the ICM, producing similar radial trends to those observed, and minimal evolution \citep{Fabjan0909.0664, Biffi1701.08164}.

Although the available data are consistent with this early enrichment scenario, constraints on the evolution of the center-excluded ICM metallicity are still relatively weak.
The only results to date that both exclude cluster centers (out to 10's of percent of $r_{500}$) and extend to redshifts $z>1$ are those of \citet{Ettori1504.02107} and \citet{Mantz1706.01476}, respectively based on XMM and \Chandra{}+{\it Suzaku} data.
For models where the metallicity at large radii evolves as $(1+z)^\gamma$, these authors respectively find $\gamma=-0.26\pm0.61$ and $\gamma=0.35^{+1.18}_{-0.36}$ from cluster samples spanning redshifts 0.09--1.4 and 0.02--1.03.
What principally limits these constraints is both the lack of data at very high redshifts, and the relatively low statistical power of the observations that do exist at $z\gtsim1$.
The results presented in Section~\ref{sec:results_thermo} represent the first constraint on the center-excluded metallicity in such a high-redshift cluster, and thus can provide significant leverage on its evolution. 
In particular, we find a metallicity of $Z/\Zsun=0.33\pm0.17$ at radii of 0.3--2.2\,$r_{500}$.

Note that the outer annulus adopted in this analysis extends to somewhat smaller radii than were used in the works cited above, $r>0.3\,r_{500}$ as compared with $>0.4\,r_{500}$ \citep{Ettori1504.02107} and 0.5--1.0\,$r_{500}$ \citep{Mantz1706.01476}.
This is motivated by the observation of \citet{Mantz1706.01476} that the metallicity in the intermediate radial range of 0.1--0.5\,$r_{500}$ is consistent with a constant at redshifts $z\gtsim0.4$, and only shows evidence of evolution when redshifts $z\ltsim0.4$ are included.\footnote{Physically, this behavior likely reflects that the central region where stellar evolution continues to enrich the ICM on average does not correspond to a neat fraction of $r_{500}$. For this reason, \citet{Liu200312426} recently advocated for a more physical core+constant model of the ICM metallicity profile, although constraining this model requires sufficient depth and spatial resolution to measure several metallicity bins per cluster. Here we continue to use a fraction of $r_{500}$ for simplicity and ease of comparison with previous work.}
This suggests that, at high redshifts, radii even as small as a few tenths of $r_{500}$ could reasonably be used to test the universality of the center-excluded metallicity.
We adopt a radius of $0.3\,r_{500}$ separating the two annuli/shells in our analysis for two reasons.
First, the extent of the PSF compared with the cluster means that further reducing the separation radius yields only minor improvements in precision for the outer shell.
Second, our most precise measurement of the center-excluded metallicity at low redshift comes from Perseus, where a spatially uniform value was measured over 76 regions extending in to $\sim0.3\,r_{500}$, despite the fact that Perseus, unlike \cl{}, hosts a well developed cool core.
For completeness, we note that our analysis yields a metallicity of $0.48\pm0.33$ at radii $>0.5\,r_{500}$ when splitting the data at that larger radius.

\begin{table*}
  \centering
  \caption{
    Global properties of \cl{}.
    Measurements are within to the characteristic radius $r_{500}$ in three dimensions in the case of mass and gas mass, and in projection for the other quantities.
    A redshift of $z=1.71$ is assumed in the derivations of radius, mass, gas mass, and luminosity.
    The impact of the redshift uncertainty on these quantities is subdominant to statistical uncertainties.
  }
  \label{tab:scaling}
  \begin{tabular}{ccccccccc}
    \hline\vspace{-3ex}\\
    $z$ & $r_{500}$ & $M_{500}$ & $\Mgas$ & $kT$ & $Z$ & $L(0. 1$--$2.4\keV)$ & $L(0.5$--$2.0\keV)$\vspace{0.3ex}\\
        & (kpc) & $(10^{14}\Msun)$ & $(10^{13}\Msun)$ & (keV) & $(\Zsun)$ & $(10^{44}\erg\second^{-1})$ & $(10^{44}\erg\second^{-1})$\vspace{0.2ex}\\
    \hline\vspace{-2ex}\\
    $1.705\pm0.018$ & $458\pm18$ & $1.8\pm0.2$ & $2.3\pm0.3$ & $7.2\pm0.3$ & $0.37\pm0.08$ & $8.4\pm0.3$ & $5.20\pm0.18$\vspace{0.5ex}\\
    \hline
  \end{tabular}
\end{table*} 

 Figure~\ref{fig:metalz} shows the {\it Suzaku} and \Chandra{} data used by \citet[][black points and boxes]{Mantz1706.01476} and the constraints on evolution models from those data alone (red shading).
Here we have cross-calibrated the \Chandra{} data to a normalization determined from {\it Suzaku}, as described by \citet{Mantz1706.01476}, since precise metallicity constraints from XMM and {\it Suzaku} in low redshift clusters agree well (e.g.\ \citealt{Leccardi0806.1445, Urban1102.2430, Urban1706.01567, Werner1310.7948, Simionescu1506.06164, Simionescu1704.01236}).
To be concrete, the model fitted is a power law, $Z = Z_0 (1+z)^\gamma$, along with a lognormal intrinsic scatter, $\sigma_{\ln Z}$.
When incorporating the center-excluded metallicity constraint from \cl{} at $z=1.71$ (blue point), we obtain $Z_0 = 0.328\pm0.015$, $\sigma_{\ln Z} < 0.06$ and $\gamma = 0.25\pm0.34$ (68.3 per cent confidence limits; blue shading in the figure).
Thanks to the high precision of the $z\approx0$ {\it Suzaku} data, and the high redshift and (relatively) good precision of our measurement for \cl{}, our constraint on $\gamma$ is tighter by a factor of $\sim2$ compared with those of \citet{Ettori1504.02107} and \citet{Mantz1706.01476}.
At the same time, there is clearly room for constraints to be improved through new, high-redshift observations, which have disproportionate leverage on the evolution parameter.

\subsection{Global scaling relations and evolution}

\begin{figure*}
  \centering
  \includegraphics{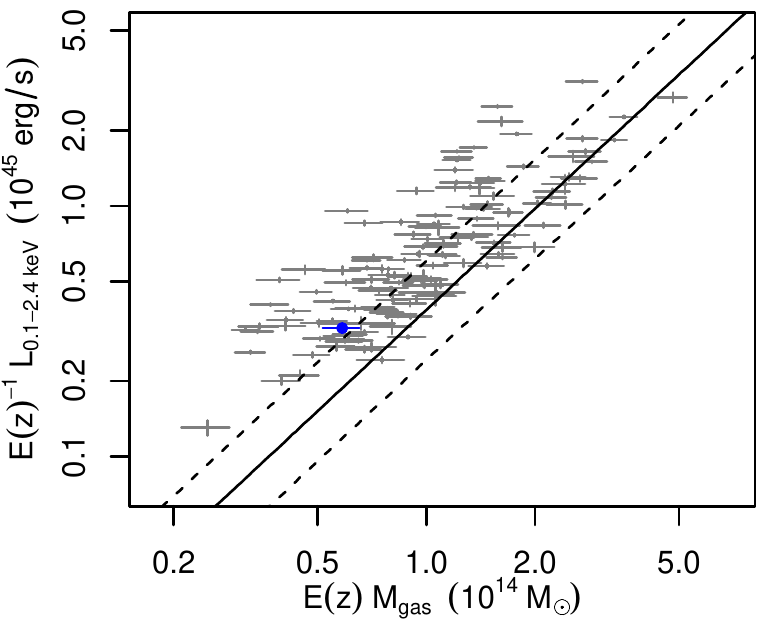}
  \hspace{8mm}
  \includegraphics{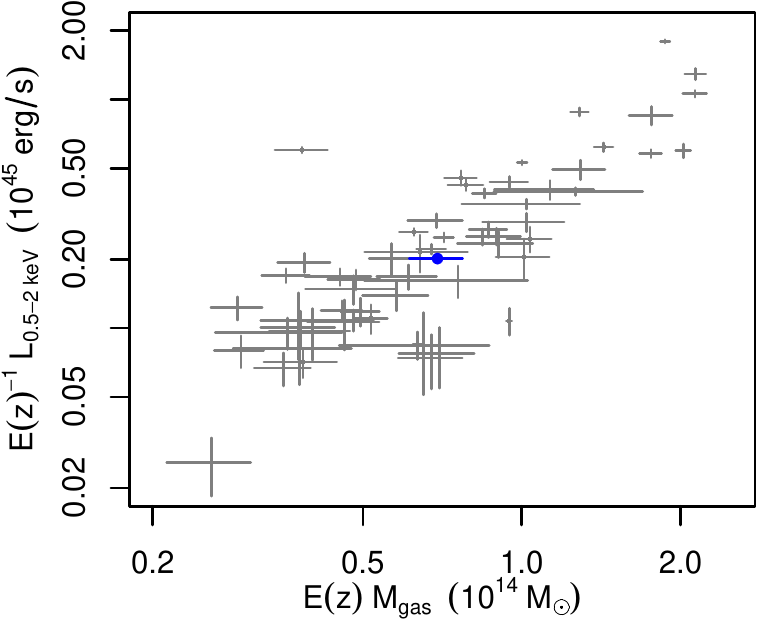}
  \vspace{3mm}\\
  \includegraphics{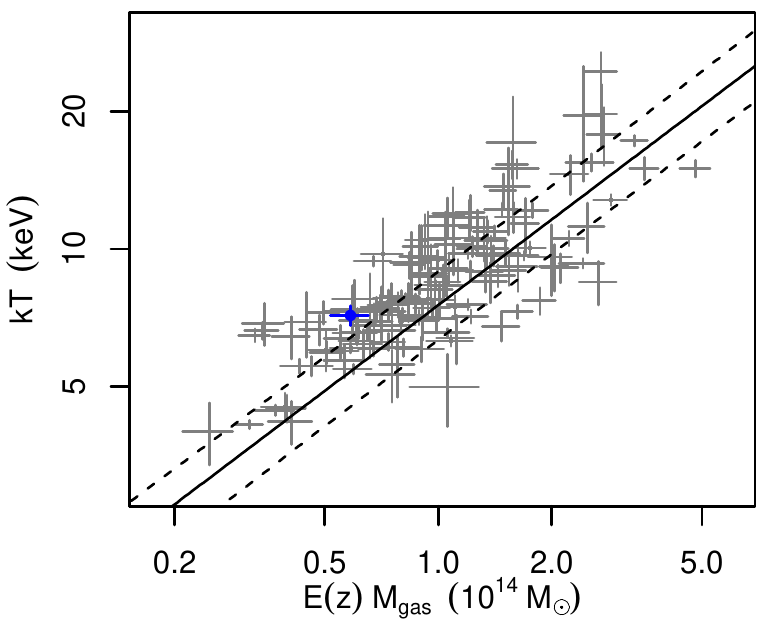}
  \hspace{8mm}
  \includegraphics{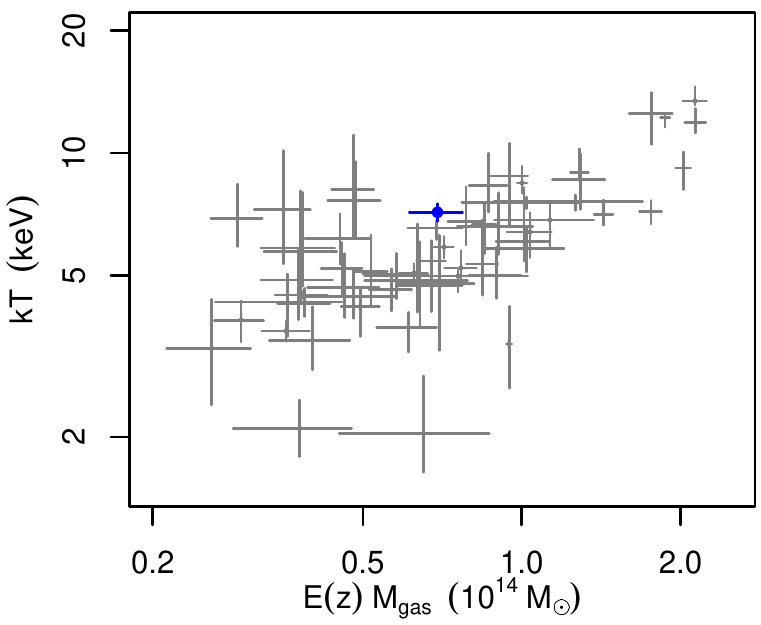}
  \vspace{3mm}\\
  \includegraphics{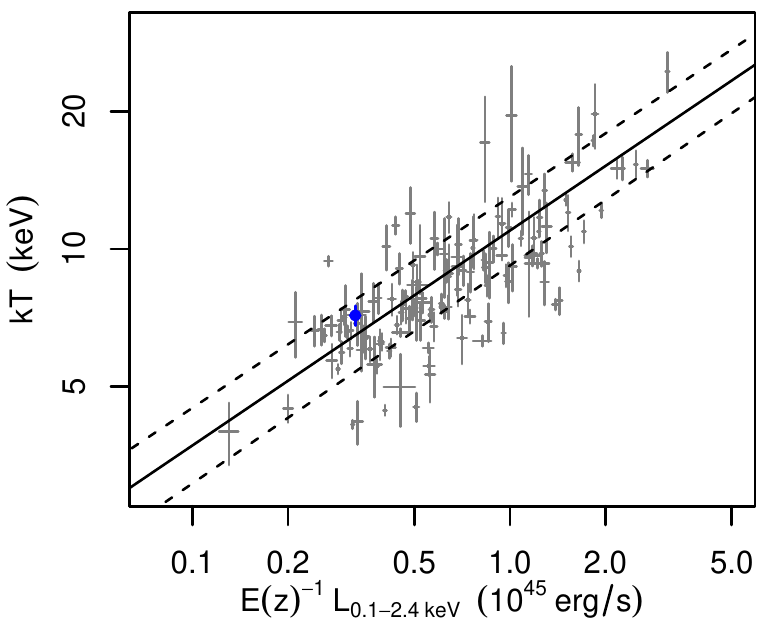}
  \hspace{8mm}
  \includegraphics{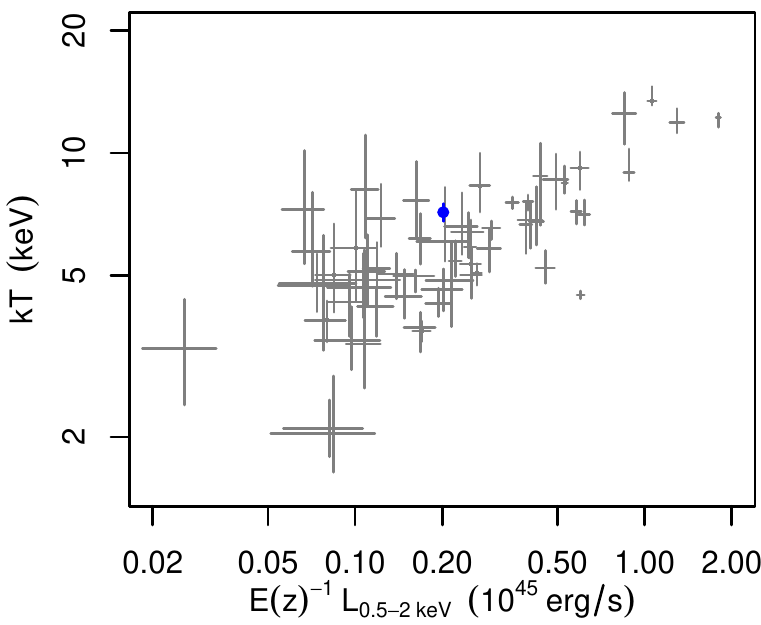}
  \caption{
    Comparison of the global properties of \cl{} (blue point) with lower-redshift cluster samples.
    Factors of $E(z)=H(z)/H_0$ encode self-similar evolution of the scaling relations.
    Left column: gray points show measurements from \citet[][median $z=0.21$]{Mantz1606.03407}, and solid/dashed lines the corresponding scaling relations (accounting for X-ray flux-selection bias) and their uncertainty (including intrinsic scatter).
    These measurements were obtained from \Chandra{} data, but the procedure for determining each observable (and $r_{500}$) is essentially the identical to the one used in this work, apart from not needing to model the PSF.
    Right column: gray points show measurements of 59 SPT clusters observed by XMM \citep[][median $z=0.44$]{Bulbul1807.02556}.
    Here the gas mass of \cl{} has been adjusted for consistency with the method used to determine $r_{500}$ for the other SPT clusters.
  } \label{fig:scaling}
\end{figure*}

Figure~\ref{fig:scaling} compares the integrated X-ray luminosity and gas mass (Section~\ref{sec:results_deproj}) and average temperature (Section~\ref{sec:results_bucket}) for \cl{} (collected in Table~\ref{tab:scaling}) with two differently selected samples of clusters at lower redshifts.
The left column compares with \Chandra{} measurements of massive, X-ray selected clusters at $0.0<z<0.5$ (median redshift 0.21) from \citet{Mantz1606.03407}, while the right column shows a subset of SPT clusters observed by XMM at $0.2<z<1.5$ (median 0.44) from \citet{Bulbul1807.02556}.
Note that, while there are well known differences between temperatures measured using \Chandra{} and XMM in general, we do not expect cross-calibration to be a significant issue for these comparisons, since \citet{Bulbul1807.02556} use only MOS (and not PN) XMM data, and the high redshift of \cl{} means that our measurements are dominated by soft observer-frame energies where the instrumental response calibrations of the telescopes are in good agreement (as verified in Section~\ref{sec:model_psf}; see also \citealt{TsujimotoAA...525A..25, Schellenberger1404.7130}).
One caveat is that the procedure for determining $r_{500}$ for the purposes of estimating these global properties used here for \cl{} is identical to that of \citet{Mantz1606.03407}, but differs from that of \citet{Bulbul1807.02556}.
We therefore applied a simple adjustment to the \Mgas{} value for \cl{} in the right column of the figure, accounting for the difference in $r_{500}$ between our determination and that of \citet{Bleem1409.0850}, as well as the slope of our measured gas mass profile. This \Mgas{} adjustment is an increase of $\sim18$ per cent, while any changes to the centrally weighted luminosity and temperature measurements should be negligible in comparison.

Since self-similar evolution in the plotted quantities \citep{Kaiser1986MNRAS.222..323K} is accounted for in Figure~\ref{fig:scaling}, we can (broadly) interpret the presence of \cl{} within the distributions of lower-redshift clusters as evidence against strong departures from self-similar evolution over the redshift ranges probed ($\sim0.2$--1.7).
This consistency with self-similar evolution is in agreement with the internal constraints on scaling relations from the two comparison samples \citep{Mantz1606.03407, McDonald1702.05094, Bulbul1807.02556}.
In comparison with both the X-ray and SZ-selected samples shown, \cl{} is perhaps relatively hot for its gas mass and luminosity, but otherwise fairly typical, given a self-similar extrapolation.
Note that we would expect this SZ-selected cluster to be relatively hot for its luminosity in comparison with the X-ray selected sample, due to the dependences of the selection observables on these physical quantities.

\section{Conclusion} \label{sec:conclusion}

We present results from a very deep XMM observation of \cl{}, the highest (confirmed) redshift SZ-discovered galaxy cluster known.
Our data provide a precise constraint on the cluster redshift of $z=1.705\pm0.018$, in agreement with earlier limits from optical and IR photometry.
Within $\sim r_{500}$, we measure an emission weighted temperature of $7.2\pm0.3$\,keV and metallicity of $Z/\Zsun=0.37\pm0.08$.
Accounting for the XMM PSF, we constrain the gas density profile, and use this to estimate the characteristic mass to be $M_{500}=(1.8\pm0.2)\E{14}\Msun$ -- equivalently, $r_{500}=(458\pm18)$\,kpc -- in our reference cosmological model.
Comparing the luminosity, temperature, and gas mass of \cl{} with lower-redshift X-ray and SZ selected cluster samples, we find broad agreement with self-similarly evolving scaling relations.

For the first time at this high redshift, the data allow us to constrain properties of the intracluster gas in the cluster center separately from the gas in the outskirts, here defined as $r<0.3\,r_{500}$ and $r>0.3\,r_{500}$.
In particular, we measure the Fe abundance in the outskirts to be $Z/\Zsun=0.33\pm0.17$, in good agreement with the ``universal'' enrichment value of 0.3 measured in the outskirts of nearby clusters.
Despite the relatively modest precision of our measurement, the high redshift of this cluster allows us to improve constraints on power-law metallicity evolution models, $\propto(1+z)^\gamma$, by a factor of 2.
Combining with lower-redshift {\it Suzaku} and \Chandra{} data, we find $\gamma=0.25\pm0.34$, consistent with no evolution in the outskirts metallicity.

That the large-scale metallicity and thermodynamic properties of \cl{} appear consistent with the simplest models of cluster formation and evolution reinforces the notion that massive, virialized clusters with deep potential wells represent (relatively) physically simple systems once they have formed.
This is good news for cluster cosmology tests using current SZ surveys with SPT-3G and Advanced ACT \citep{Benson1407.2973, Henderson1510.02809}, as well as upcoming SZ and X-ray surveys with the Simons Observatory, CMB-S4 and ATHENA \citep{Nandra1306.2307, CMBS4-Science-Book, Cucchetti1809.08903, Ade1808.07445}, which will find clusters to redshifts $z>2$.
To place these observations on a firm quantitative footing, however, will require dedicated X-ray follow-up of clusters found at these redshifts; for metallicity studies in particular, this requires deeper data than that normally obtained to estimate simple mass proxies.
Ultimately, this work at $z>2$ will benefit from next-generation observatories like ATHENA and {\it Lynx}, which are planned for launch in the 2030's.
In the interim, at slightly lower redshifts, there remains a clear role for deep observations with XMM, such as the one used here.

\section*{Acknowledgments}

We thank Stefano Ettori for helpful comments.

We acknowledge support from the National Aeronautics and Space Administration under Grant No.\ 80NSSC18K0578, issued through the XMM-{\it Newton} Guest Observer Facility; and from the U.S.\ Department of Energy under contract number DE-AC02-76SF00515.  
 
This work was performed in the context of the South Pole Telescope scientific program. SPT is supported by the National Science Foundation through grant PLR-1248097. Partial support is also provided by the NSF Physics Frontier Center grant PHY-0114422 to the Kavli Institute of Cosmological Physics at the University of Chicago, the Kavli Foundation and the Gordon and Betty Moore Foundation grant GBMF 947 to the University of Chicago.
The SPT is also supported by the U.S.\ Department of Energy.
Work at Argonne National Lab is supported by UChicago Argonne LLC, Operator of Argonne National Laboratory (Argonne). Argonne, a U.S.\ Department of Energy Office of Science Laboratory, is operated under contract no. DE-AC02-06CH11357.

\def \araa {ARA\&A}
\def \aj {AJ}
\def \aar {A\&AR}
\def \apj {ApJ}
\def \apjl {ApJL}
\def \apjs {ApJS}
\def \asl {Adv. Sci. Lett.} 
\def \mnras {MNRAS}
\def \nat {Nat}
\def \pasj {PASJ}
\def \pasp {PASP}
\def \science {Sci}
\def \gca {Geochim.\ Cosmochim.\ Acta}
\def \npa {Nucl.\ Phys.\ A}
\def \plb {Phys.\ Lett.\ B}
\def \prc {Phys.\ Rev.\ C}
\def \prd {Phys.\ Rev.\ D}
\def \prl {Phys.\ Rev.\ Lett.}
\def \jcap {J. Cosmology Astropart. Phys.} 
\def \physrep {Phys. Rep.} 
\def \aap {A\&A} 
\def \ijmpd {Int.\ J.\ Mod.\ Phys.\ D} 

\end{document}